
%
%
%
%
\magnification=\magstep1
\baselineskip=7truemm plus 1truemm minus 0.1truemm
\def\gap{\;_\sim^>\;}
\def\lap{\;_\sim^<\;}
\line {\hfil IASSNS-HEP-94/39, PUPT-1470}
\line {\hfil POP-568, CfPA-TH-94-27, UTAP-185}
\line {\hfil June 1994}
\centerline{CBR ANISOTROPY IN AN OPEN INFLATION, CDM COSMOGONY}
\bigskip
\centerline{Marc Kamionkowski$^1$, Bharat Ratra$^2$, David N. Spergel$^3$,
and Naoshi Sugiyama$^{4,5}$}
\centerline{$^1\!${\it School of Natural Sciences, Institute for
Advanced Study, Princeton, NJ 08540}}
\centerline{$^2\!${\it Joseph Henry Laboratories, Princeton University,
Princeton, NJ 08544}}
\centerline{$^3\!${\it Princeton University Observatory, Princeton,
NJ 08544}}
\centerline{$^4\!${\it Department of Astronomy, University of California,
Berkeley, CA 94720}}
\centerline{$^5\!${\it Department of Physics, Faculty of Science,
University of Tokyo, Tokyo 113}}
\bigskip
\centerline{ABSTRACT}
\smallskip

We compute the cosmic background radiation anisotropy,
produced by energy-density fluctuations generated during
an early epoch of inflation, in an open cosmological model
based on the cold dark matter scenario. At $\Omega_0 \sim$
0.3 --  0.4, the COBE normalized open model appears to be consistent
with most observations.

\bigskip
\noindent  {{\it Subject headings:} cosmic microwave background --- large-scale
structure of the universe --- galaxies: formation}
\bigskip
\centerline {Submitted to {\it Astrophysical Journal Letters}}


\bigskip
\centerline{1. INTRODUCTION}
\medskip

Observational evidence (summarized in Peebles 1993 and
Ratra \&\ Peebles 1994b, hereafter RPb) suggests that the cosmological
clustered mass density parameter, $\Omega_0$, is significantly
smaller than the Einstein-de Sitter value of unity, but possibly
somewhat larger than the baryon density value predicted from
the standard nucleosynthesis consideration (Walker et al. 1991).
Among the low-density cold dark matter (CDM) cosmogonies now under discussion,
a model consistent with the familiar version of the
inflation picture (Guth 1981; Kazanas 1980; Sato 1981a,b)
is a low-density flat universe dominated by a cosmological
constant $\Lambda$ (Efstathiou, Sutherland, \& Maddox 1990;
Kofman, Gnedin, \& Bahcall 1993; Stompor \& G\'orski 1994),
while a low-density model with open spatial hypersurfaces and
$\Lambda = 0$ (Ratra \&\ Peebles 1994a, hereafter RPa; RPb,
and references therein) could perhaps be accommodated in a
variant of the inflation picture in which a single-bubble
open inflation model is created by tunnelling in a spatially-flat
de Sitter spacetime which also inflates (Gott 1982; Guth \&\ Weinberg 1983).
In the open case, the first epoch of inflation smooths away
initial inhomogeneities, which, if significant on the scale set by
space curvature in the second epoch of inflation, would result in an
unacceptable large-scale CBR anisotropy (Kashlinsky, Tkachev, \&\ Frieman
1994).

In a model with open spatial sections, the radius of curvature of the space
sections introduces a new global length scale (in addition to that
set by the Hubble parameter, $H$), and one can
either assume a simple functional form for the spectrum of energy
density perturbations (Wilson 1983; Sugiyama \&\ Gouda 1992;
Kamionkowski \&\ Spergel 1993, hereafter KS; Sugiyama \&\ Silk 1994,
hereafter SS), or
compute the spectrum that arises from quantum-mechanical
zero-point fluctuations during an early epoch of inflation in
an open model (Lyth \&\ Stewart 1990; Ratra 1994; RPa).

In RPb the spectrum that results from such a
computation, and a generalization to the open
model of the Sachs-Wolfe relation between the cosmic
background radiation (CBR) anisotropy and the mass distribution
(Anile \&\ Motta 1976; RPa), were used to determine the CBR
quadrupole anisotropy, $Q$. To fix the inflation-epoch
parameters of the model $Q = 10 e^{\pm 1} \mu$K was taken
as the range allowed by the measurements (Bennett et al. 1994, hereafter B94;
Wright et al. 1994b; Ganga et al. 1994; G\'orski et al. 1994).
A number of statistics of cosmological
interest were then estimated, with results that were observationally
encouraging,
but with large uncertainty because of the relatively large range
of $Q$ allowed by the observations and by theoretical cosmic
variance. SS have recently studied large-scale CBR anisotropies
in this and other low-density models; they, however, did not
examine large-scale structure.

Here we summarize a computation of the lowest two thousand
CBR multipoles in this model,
use the result to normalize the model to the anisotropy
at $10^\circ$ (which is observationally better determined
than $Q$, and has smaller cosmic variance), and tabulate
statistics of cosmological interest.
In agreement with earlier conclusions, depending on the
$10^\circ$ CBR anisotropy, when
$\Omega_0 \sim 0.3$ or maybe somewhat larger, but still significantly
below the Einstein-de Sitter value, the open
model does fairly well at fitting most observations.
In addition, the shape of the large-scale CBR anisotropy
multipole spectrum differs from that in the $\Omega_0 = 1$ CDM model,
and may allow an observational test of this open model.

The open inflation model is discussed in \S 2. In \S 3
we summarize the CBR anisotropy computation, and
in \S 4 we consider the predictions of the model.

\bigskip

\centerline{2. MODEL AND POWER SPECTRUM}\nobreak
\medskip\par

The inflation epoch of the open cosmological model of
RPa,b is characterized by the potential for the inflaton
scalar field $\Phi$,
$$
    V(\Phi) = 12 \bar h^2 \left[ 1 - \epsilon \Phi \right] ,
    \eqno(1)
$$
where the first term, $12\bar h^2$, is responsible for the expansion during
inflation, and the second term, with $\epsilon$ small, forces
the mean value of $\Phi$ `down the hill'. At reheating
$V(\Phi)$ vanishes and the $\Phi$ energy density
is converted to radiation energy density (Ratra 1992).

The computation of the fluctuations produced during inflation
is described in RPa and the results are summarized in RPb.
We work to linear order in the matter and metric perturbations
about a spatially homogeneous open cosmological model, and in the inflation
epoch we also work to lowest nontrivial order in an
expansion in $\epsilon$ (Ratra 1989).

One approach to computing the CBR anisotropy makes use
of the gauge-invariant fractional energy-density perturbation
$(\Delta )$ power spectrum, $P_\Delta (A, t) = |\Delta (A, t)|^2$,
where the radial coordinate wavenumber $A \, (0 < A < \infty )$
is related to the eigenvalue of the spatial scalar Laplacian,
$ -(A^2 + 1)$. ($P_\Delta$ should not be confused with the
instantaneously Newtonian synchronous hypersurface power spectrum
used in RPb.) The present linear-theory power spectrum is (RPa)
$$
   {\epsilon^2 \over (1+z_{\rm re})^4} P_\Delta (A)
   = 2 \pi \left({H_0 \over m_p}\right)^2 {\Omega_0
   \over 1+z_{\rm eq}} \left({W_1 \over c_1}\right)^2
   { (4+A^2)^2 \over A(1+A^2)} ,
   \eqno(2)
$$
where the Planck mass $m_p = G^{-1/2}$, $H_0 = 100 h$ km s$^{-1}$
Mpc$^{-1}$ is the present Hubble parameter, $z_{\rm eq}$ is the
redshift of equality of radiation and matter mass densities,
$z_{\rm re}$ that of reheating, and
$$
   {W_1\over c_1} = {2(1-\Omega_0) \over 45 \Omega_0{}^{1.5}
   (1+z_{\rm eq})^{2.5}} + {1+2\Omega_0 \over 1-\Omega_0}
   + {3\Omega_0 \over (1-\Omega_0)^{1.5}} {\rm ln}
   \left\{ {1\over \sqrt{\Omega_0}} - \sqrt{1-\Omega_0\over \Omega_0}
   \right\} .
   \eqno(3)
$$
The first term on the right hand side is a non-power-law correction
to the adiabatic solution; it is subdominant unless $\Omega_0$
is very small. The wavenumber dependence of equation (2) is
consistent with that of the energy-density perturbation power
spectrum which may be derived from the expressions of Lyth \&\ Stewart
(1990) (Lyth 1994). As noted in RPa,b, at short wavelengths
$|\Delta|^2 \propto A$ (the usual $n = 1$ scale-invariant form),
while at long wavelengths $|\Delta|^2 \propto 1/A$ (this growth
at small $A$ is not disturbing,
since on large scales the spatial harmonics are strongly damped).
Finally, we note that $P_\Delta \propto 1/\epsilon^2$, so an
observational upper bound on $P_\Delta$ results in a lower limit
on $\epsilon$, the slope of the inflaton potential (Ratra 1992,
1989, 1990).

\bigskip
\centerline{3. CBR ANISOTROPY}\nobreak
\medskip\par

The computation of the CBR anisotropy multipole moments
$C_l = \langle |a_l^m|^2 \rangle$ (where the temperature
anisotropy in comoving coordinates is $\delta T/T =
\sum_{l,m} a_l^m Y_l^m$) makes use of the gauge-invariant
formalism of Gouda, Sugiyama, \&\ Sasaki (1991). (It can be
shown that this is identical to the synchronous-gauge formalism
of RPa.) In this preliminary computation of the $C_l$ we take
$h = 0.5$ and the present baryon density $\Omega_B = 0.03$, use
the wavenumber dependence of the power spectrum in equation
(2), numerically integrate the perturbation equations
starting from well before $z_{\rm eq}$ (as a result
we may ignore the subdominant non-power-law term in eq. [3]),
and account for the fuzziness of the last-scattering surface
(the CBR anisotropy on this surface is negligible). We
find that the scaled quadrupole $(l=2)$ moment ($Q^2$ multiplied
by $\epsilon^2/(1+z_{\rm re})^4$, as in eq. [2]), agrees
with that found in RPb to better than 1\% (we compared the
two computations of the ratios of $Q$ at total $\Omega_0
= 0.1, 0.2,$ and $0.2, 0.3, \cdots ,$ and 0.5, 0.6;
since we use the numerical values from the $h = 0.8$ and $\Omega_B = 0$
run of the RPb computations, this result shows that for all
practical purposes $Q$ is independent of $h$ and $\Omega_B$,
Bond et al. 1994).

Using the numerical values for the $C_l$, one could fix the model
normalization by requiring that the rms temperature anisotropy
at $10^\circ$ angular resolution agree with the two year COBE
value $\delta T = 30.5 (1\pm 0.16) \mu$K (B94),
where the range is that allowed at one standard deviation from
the measurement errors and model-dependent cosmic variance added
in quadrature. This $\delta T$ is determined from
the data after a monopole and dipole is subtracted (which affects
the value of the quadrupole and octupole); as a result the cosmic
$\delta T$ is likely to be somewhat larger than $30.5 \mu$K
(Wright et al. 1994a). Also, if one uses, as we do, a $10^\circ$ FWHM
gaussian approximation to the DMR beam shape, one must
increase $\delta T$ (Wright et al. 1994a). It does not
yet seem possible to account for these adjustments in a (theoretical)
model independent manner, but for the purpose of this preliminary
comparison it suffices to adopt $\delta T(10^\circ) = 35 (1 \pm 0.3) \mu$K,
where we have taken the precaution of increasing the range to
account for possible model-dependent effects. On this large
a scale $(10^\circ)$ the dependence of $C_l$ on $h$ and $\Omega_B$ is
very weak,
and so the normalization is almost independent of the value of
$h$ and $\Omega_B$. The numbers in column (2) of the table is the
value of $Q$ predicted with this normalization.
Comparing to the result of RPb, we see that the parameters of
the inflation epoch model must obey $1 + z_{\rm re} \sim 10^{29}
\sqrt{\epsilon}$.

The numbers in columns (3) and (4) of the table are the present
rms linear fluctuation $\delta M/M$ in the mass averaged over a sphere of
radius
8$h^{-1}$ Mpc and the present rms value $v_p$ of the line-of-sight
peculiar velocity in a window of radius 50$h^{-1}$ Mpc. They
are scaled by the ratio of $Q$ estimated here to that of RPb,
{}from the numerical values computed in RPb. (RPb took $\Omega_B = 0$,
so the $\delta M/M$ numbers in the table are fair for $\Omega_0 \sim
0.3$ and larger, while the rough estimate of $v_p$ in RPb assumed that
on 50$h^{-1}$ Mpc the transfer function could be ignored.)
The range in each entry only accounts for the uncertainty in
the $10^\circ$ normalization.

In Figures 1 and 2 we show the $C_l$, to $l = 50$
and $l = 2000$, as a function of $l$. The highest curve in the
figures, at $ l \sim 50$, is the flat model, with $\Omega_0 = 1$, and the
other curves are the open models, all with $h = 0.5$
and $\Omega_B = 0.03$. For $l < 50$ the dependence on $h$,
$\Omega_B$, and ionization history is very weak, but for larger $l$ the
dependence is significant, and so the $l > 100$ part of
the curves in Figure 2 are only meant to be illustrative.
Finally, we show in Figure 3 the scaled Newtonian hypersurface
physical linear energy density perturbation power spectrum,
$(a_0 h)^3 \widehat P(A) T^2(A)$ (RPb, eq. [2],
where $a_0$ is the present cosmological scale factor,
and $T^2(A)$ is the transfer function with $\Omega_B = 0$),
as a function of the scaled proper wavenumber, $A/(a_0 h)$.
The highest curve is the Einstein-de Sitter
model, with $\Omega_0 = 1$ and $h = 0.5$; the other curves
are the open models with $h = 0.65$.

\bigskip
\centerline{4. DISCUSSION}\nobreak
\medskip\par\nobreak

Dynamical mass estimates on length scales $\lap 10 h^{-1}$Mpc
consistently suggest $\Omega_0 = 0.2 \pm 0.1$
(Peebles 1993), while the preliminary dynamical evidence,
on scales $\gap 20 h^{-1}$Mpc, from the IRAS/POTENT analysis
of large-scale flows is that $\Omega_0 \approx 1$ (Dekel et al. 1993).
Large-scale estimates based purely on redshift surveys, however,
are consistent with lower $\Omega_0$ (Hamilton 1993; Fisher et al. 1994).
Also, as summarized in RPb, most
of the rest of the observational evidence is consistent
with a low-density open or $\Lambda$-dominated flat model.
In our computation of $\delta M/M$ and $v_p$ we adopt $h = 0.65$
(when $\Omega_0 < 1$), which is in the
range of most recent estimates (Jacoby et al. 1992; van den Bergh 1992;
Fukugita, Hogan, \& Peebles 1993; Birkinshaw \& Hughes 1994;
Sandage et al. 1994; Schmidt et al. 1994). For $\Omega_0 = 0.4$
this implies an expansion time $\sim 12$ Gyr, which is consistent
with, but near the low end of, recent estimates (van den Bergh 1992).

The rms fluctuation in the number of galaxies in a randomly
place sphere of radius $8h^{-1}$Mpc is observed to be
$\delta N/N =$ 0.79 to 1.1 (Peebles 1993, eqs. [7.33,
7.73]). From Table 1 we see that, depending on the
$10^\circ$ CBR anisotropy, when $\Omega_0 = 0.3$ the open model
could be consistent with a bias factor of about two.
An $\Omega_0 = 0.1$ model would require unreasonably high bias, while
$\Omega_0 > 0.4$ could be
consistent with no bias. Given the uncertainties,
these values of the bias factor
should also suffice to make the spectra of Figure 3 consistent
with the data. (The $P_\Delta (A) \propto A$ spectrum considered
by KS has less large-scale power than the one considered
here, and when normalized to COBE it results in a $\delta M/M(8
h^{-1} {\rm Mpc})$ that is $\sim 25\%$ larger at $\Omega_0 = 0.3$.)
The observed cluster mass and correlation functions provide another test
(Lilje 1992; Kauffmann \& White 1992; Weinberg \& Cole 1992;
Oukbir \& Blanchard 1992; Bahcall \& Cen 1992;
White, Efstathiou, \& Frenk 1993). For instance, Cen, Gnedin,
\& Ostriker (1993) find that an $\Omega_0 h = 0.2$, $\Lambda$-dominated
$\Omega_0 = 0.3$ model, with $\delta M/M(8 h^{-1} {\rm Mpc}) = 0.67$,
is a reasonable fit to the data. From Table 1 we see that in the
open model at $\Omega_0 = 0.4$, $\Omega_0 h = 0.26$ and,
depending on the $10^\circ$ CBR anisotropy,
$\delta M/M(8 h^{-1} {\rm Mpc}) \sim 0.6$.

It is interesting that the values predicted for $Q$ after
normalizing at $10^\circ$, column (2) of the table, are
larger than the COBE CBR measurement $6(1\pm 0.5) \mu K$ (B94).
This is also the case in the Einstein-de Sitter model,
but not for topological defects in an open universe (Spergel 1993).
Since the total quadrupole is significantly
affected by emission from our galaxy, and cosmic variance is
non-negligible, it would be premature to conclude that this rules
out the model. (B94 note that the probability of finding the
one-standard-deviation measured COBE range from a flat model
with $Q = 17\mu$K is 10\%.)
The shape of the low-order CBR multipoles (Fig. 1) is
quite insensitive to the value of $h$ and $\Omega_B$, but
does depend on the value of $\Omega_0$. The shape of the
low-$\Omega_0$ open inflation model spectrum is somewhat
reminiscent of that in the scale-invariant
$\Lambda$-dominated flat model and in the tilted CDM model,
but differs from that of the scale-invariant Einstein-de Sitter
case (SS). The shape at low $l$
is mostly determined by two effects: the strong damping of the
open model spatial harmonics on scales comparable to that of space
curvature; and the long wavelength $1/A$ form of the
power spectrum (eq. [2]). Relative to the $P_\Delta(A) \propto A$ model
(where the shape of the low-$l$ $C_l$ is determined by the
damping, KS), we see that here the asymptotic $1/A$ behaviour
opposes the damping and raises the low-$l$
$C_l$, as long as $\Omega_0$ is not too small (the
$\Omega_0 = 0.1$ multipoles at $l =$3 -- 5 are larger than
at $l = 2$; this might be because the present Hubble scale is closer
to the space curvature scale and so the damping is
more significant for $l = 2$, which goes out to larger scales.)
It would be useful to more carefully compare the
low-$l$ $C_l$ to the data (and thereby
more accurately fix the model normalization).
The large $l$ part of the spectrum
(Fig. 2) is much more sensitive to the ionization history and
the values of $h$ and $\Omega_B$. We see, as
noted by Kamionkowski et al. (1994), that
in the open case the position of the peak in the spectrum is sensitive to the
value of $\Omega_0$, but insensitive to the large
wavenumber form of $P_\Delta(A)$, and depends weakly on $\Lambda$,
$\Omega_B$, $h$, and ionization history. Observations of small-scale
CBR anisotropies thus might allow for a discrimination between
$\Lambda$-dominated and open models.

Finally, we emphasize that structure formation occurs earlier
in the low-density open and $\Lambda$-dominated flat CDM models,
compared to the $\Omega_0 = 1$ tilted CDM and mixed dark matter
cases (RPb). It would be of some interest to more carefully
quantify the differences, since with moderately high redshift
data one should be able to see the significant evolution of large-scale
structure predicted in those models in which structure forms late.

\bigskip
We thank D. Bond, K. Ganga, K. G\'orski, R. Gott, L. Page,
J. Silk, N. Turok, D. Weinberg,
E. Wright, and especially J. Peebles for helpful discussions.
This work was supported in part by the Department of Energy under
contract DEFG02-90-ER40542, by NSF grants PHY89-21378, AST88-58145
and ASC93-18185,
by the David and Lucile Packard Foundation, by NASA grant NAGW-2448,
and by a JSPS Postdoctoral Fellowship for Research Abroad.

\vfill\eject

\centerline{TABLE 1}
\centerline{Numerical Values${}^{\rm a}$}
\bigskip
\hbox to \hsize{\hss
\vbox{
\hrule height 0.6pt
\vskip 2pt
\hrule height 0.6pt
\halign{\strut
        \ #\quad\hfil & \hfil\quad#\quad\hfil
	& \hfil\quad#\quad\hfil & \hfil\quad#\quad\hfil \cr
\noalign{\vskip 3pt}
$\Omega_0$ & $Q^{\rm b}$ &
${\delta M \over M}(8h^{-1}{\rm Mpc})$ & $v_p(50h^{-1}{\rm Mpc})^{\rm c}$ \cr
(1) & (2) & (3) & (4) \cr
\noalign{\vskip 3pt}
\noalign{\hrule height 0.6pt}
\noalign{\vskip 3pt}
0.1 & $18(1\pm 0.3)$ & 0.046 -- 0.084 & 62 -- 120 \cr
0.2 & $20(1\pm 0.3)$ & 0.13 -- 0.25 & 100 -- 190 \cr
0.3 & $20(1\pm 0.3)$ & 0.25 -- 0.46 & 140 -- 260 \cr
0.4 & $19(1\pm 0.3)$ & 0.39 -- 0.72 & 180 -- 340 \cr
0.5 & $17(1\pm 0.3)$ & 0.54 -- 1.0 & 220 -- 410 \cr
1 & $17(1 \pm 0.3)^{\rm d}$ & 0.73 -- 1.4${}^{\rm d}$
& 280 -- 520${}^{\rm d}$ \cr
\noalign{\vskip 3pt}
\noalign{\hrule height 0.6pt}
\noalign{\vskip 3pt}
}}
\hss}
${}^{\rm a}h = 0.65$ unless otherwise indicated
${}^{\rm b}$unit = $\mu$K
${}^{\rm c}$unit = km s$^{-1}$
${}^{\rm d}h = 0.5$ flat CDM model
\vfill\eject

\def\ref{\hangindent=5ex\hangafter=1}
\parskip=0pt
\parindent=0pt
\centerline{REFERENCES}
\medskip

\ref Anile, A. M., \&\ Motta, S. 1976, ApJ, 207, 685

\ref Bahcall, N. A., \&\ Cen, R. 1992, ApJ, 398, L81

\ref Bennett, C. L. et al. 1994, COBE preprint 94-01 (B94)

\ref Birkinshaw, M., \&\ Hughes, J. P. 1994, ApJ, 420, 33

\ref Bond, J. R. 1994, CITA preprint CITA-94-5

\ref Bond, J. R. et al. 1994, Phys. Rev. Lett., 72, 13

\ref Cen, R., Gnedin, N. Y., \&\ Ostriker, J. P. 1993, ApJ, 417, 387

\ref Cheng, E. S. et al. 1994, ApJ, 422, L37

\ref de Bernardis, P. et al. 1994, ApJ, 422, L33

\ref Dekel, A. et al. 1993, ApJ, 412, 1

\ref Dragovan, M. et al. 1994, ApJ, 427, L67

\ref Efstathiou, G., Sutherland, W. J., \&\ Maddox, S. J. 1990,
Nature, 348, 705

\ref Fisher, K. B. et al. 1993, ApJ, 402, 42

\ref Fisher, K. B. et al. 1994, MNRAS, 267, 927

\ref Fukugita, M., Hogan, C. J., \&\ Peebles, P. J. E. 1993, Nature, 366, 309

\ref Ganga, K., Page, L., Cheng, E., \&\ Meyer, S. 1994, Princeton preprint

\ref G\'orski, K. M. et al. 1994, COBE preprint 94-08

\ref Gott III, J. R. 1982, Nature, 295, 304

\ref Gouda, N., Sugiyama, N., \&\ Sasaki, M. 1991, Prog. Theo. Phys.,
85, 1023

\ref Guth, A. 1981, Phys. Rev., D23, 347

\ref Guth, A., \&\ Weinberg, E. J. 1983, Nucl. Phys., B212, 321

\ref Hamilton, A. J. S. 1993, ApJ, 406, L47

\ref Hancock, S. et al. 1994, Nature, 367, 333

\ref Jacoby, G. H. et al. 1992, PASP, 104, 599

\ref Kamionkowski, M., \&\ Spergel, D. N. 1993, IAS preprint
IASSNS-HEP-93/73 (KS)

\ref Kamionkowski, M., Spergel, D. N., \&\ Sugiyama, N. 1994, ApJ, 426, L57

\ref Kashlinsky, A., Tkachev, I. I., \&\ Frieman, J. 1994, Fermilab
preprint FERMILAB-Pub-94/114-A

\ref Kauffmann, G., \&\ White, S. D. M. 1992, MNRAS, 258, 511

\ref Kazanas, D. 1980, ApJ, 241, L59

\ref Kofman, L. A., Gnedin, N. Y., \&\ Bahcall, N. A. 1993,
ApJ, 413, 1

\ref Lilje, P. B. 1992, ApJ, 386, L33

\ref Lyth, D. H. 1994, private communication

\ref Lyth, D. H., \&\ Stewart, E. D. 1990, Phys. Lett., B252, 336

\ref Meinhold, P. et al. 1993, ApJ, 409, L1

\ref Oukbir, J., \&\ Blanchard, A. 1992, A\&{}A, 262, L21

\ref Peebles, P. J. E. 1993, Principles of Physical Cosmology
(Princeton: Princeton University Press)

\ref Ratra, B. 1989, Caltech preprint GRP-217/CALT-68-1594

\ref Ratra, B. 1990, Caltech preprint GRP-229/CALT-68-1666

\ref Ratra, B. 1992, Phys. Rev., D45, 1913

\ref Ratra, B. 1994, Princeton preprint PUPT-1443

\ref Ratra, B., \&\ Peebles, P. J. E. 1994a, Princeton preprint
PUPT-1444 (RPa)

\ref Ratra, B., \&\ Peebles, P. J. E. 1994b, Princeton preprint
PUPT-1445 (RPb)

\ref Readhead, A. C. S. et al. 1989, ApJ, 346, 566

\ref Sandage, A. et al. 1994, ApJ, 423, L13

\ref Sato, K. 1981a, Phys. Lett., 99B, 66

\ref Sato, K. 1981b, MNRAS, 195, 467

\ref Schmidt, B. P. et al. 1994, CfA preprint 3794

\ref Schuster, J. et al. 1993, ApJ, 412, L47

\ref Spergel, D. N. 1993, ApJ, 412, L5

\ref Stompor, R., \&\ G\'orski, K. M. 1994, ApJ, 422, L41

\ref Sugiyama, N., \&\ Gouda, N. 1992, Prog. Theo. Phys., 88, 803

\ref Sugiyama, N., \&\ Silk, J. 1994, Phys. Rev. Lett., in press (SS)

\ref Tucker, G. S., Griffin, G. S., Nguyen, H. T., \&\ Peterson, J. B.
1993, ApJ, 419, L45

\ref van den Bergh, S. 1992, PASP, 104, 861

\ref Walker, T. P. et al. 1991, ApJ, 376, 51

\ref Weinberg, D. H., \&\ Cole, S. 1992, MNRAS, 259, 652

\ref White, S. D. M., Efstathiou, G., \&\ Frenk, C. S. 1993,
MNRAS, 262, 1023

\ref Wilson, M. L. 1983, ApJ, 273, 2

\ref Wollack, E. J. et al. 1993, ApJ, 419, L49

\ref Wright, E. L. et al. 1994a, ApJ, 420, 1

\ref Wright, E. L., Smoot, G. F., Bennett, C. L., \&\ Lubin, P. M. 1994b,
COBE preprint 94-02


\vfill\eject

\centerline{FIGURE CAPTIONS}

Fig. 1.-- CBR anisotropy multipole moments $l(l+1)C_l/(2\pi)\times 10^{10}$
as a function of $l$, to $l=50$, for the open model
with $\Omega_0 = 0.1, 0.2, 0.3, 0.4,$ and 0.5, and the
Einstein-de Sitter model with $\Omega_0 = 1$, all for $h = 0.5$
and baryon density $\Omega_B = 0.03$, normalized to an
rms temperature anisotropy of 35$\mu$K at $10^\circ$.
The curves are in descending
order of $\Omega_0$ as one moves down the right hand side of the
figure. From left to right, the data points (courtesy of L. Page,
{}from Bond 1994), with vertical one
standard deviation error bars and horizontal bars centered on
the relevant window function maxima that give the
value of $l$ at which the window function falls to
$e^{-0.5}$ of the maxima, are the
flat-model power-law-spectrum multipole fit from COBE (B94),
FIRS (Ganga et al. 1994), and the lower end of the error
bar from Tenerife (Hancock et al. 1994). We emphasize that these are
preliminary estimates.

Fig. 2.-- CBR anisotropy multipoles to $l = 2000$. Aside from
the different scales on the axes, the notation is the
same as in Figure 1. The alternating solid and dashed lines
are the spectra of the open
model with $\Omega_0 = 0.1, 0.2, 0.3, 0.4,$ and 0.5, and
the Einstein-de Sitter model, as one moves up the figure
at $l \sim 100$. From left to right,
the data points (courtesy of L. Page, from Bond 1994) are
the flat-model power-law-spectrum
multipole fit from COBE, FIRS, Tenerife, ACME
(Schuster et al. 1993), Saskatoon (Wollack et al. 1993), the
lower end of the Python error bar (with most likely value
$\sim 6.7$ and half-power points $l \sim 52$ and 200, Dragovan et al. 1994),
ARGO (de Bernardis et al. 1994), MSAM2 with and without sources
(Cheng et al. 1994), MAX-MuPeg (Meinhold et al. 1993; not shown is MAX-GUM
with lower bound $\sim 6.2$),
MSAM3 with and without sources (Cheng et al. 1994), and,
with no vertical error bars, the two standard deviation
upper limits from WD (95\% CL upper limit, Tucker et al. 1993)
and OVRO (97.5\% Bayesian probability, Readhead et al. 1989).
We emphasize that these are preliminary estimates. For $l \gap 100$
the spectra are sensitive to the assumed values of $h$ and
$\Omega_B$, so this part of the figure is only meant to be
illustrative. In particular, increasing $h$ from 0.5 to 0.65
should allow the low-density open models to comply with the WD
and OVRO constraints; this could also be accomplished by early
mild reionization (Kamionkowski, Spergel, \&\ Sugiyama 1994).

Fig. 3.--  Newtonian hypersurface (scaled) physical linear power spectrum of
fractional energy density
perturbations, $P(k) [= a_0{}^3 \widehat P(A) T^2(A)]$, at the present
epoch, as a function of (scaled)
proper wavenumber $k(=A/a_0)$. The highest curve is the flat
CDM model with $\Omega_0 = 1$, $\Omega_B = 0$, and $h = 0.5$; with the
$10^\circ$ CBR
normalization $\delta M/M (8 h^{-1} {\rm Mpc}) = 1.0$.
The other curves are the
open models with $h = 0.65$, $\Omega_B = 0$, and $\Omega_0 = 0.5, 0.4, 0.3,
0.2, 0.1$ as one moves down the right hand side of the
figure. With the $10^\circ$ CBR normalization $\delta M/M (8 h^{-1} {\rm Mpc})
= 0.36$ (when $\Omega_0 = 0.3$), $=0.56 (\Omega_0 = 0.4)$, and
$=0.77 (\Omega_0 = 0.5)$, larger than the values used
in RPb. The points are the IRAS 1.2Jy redshift data
rescaled to real space under the assumptions that $\Omega_0 = 1$
and that IRAS galaxies are unbiased. They are estimated from Figure 10
of Fisher et al. (1993).

\bye